# Data-parallel programming with Intel Array Building Blocks (ArBB)

Volker Weinberg [*]

*Leibniz Rechenzentrum der Bayerischen Akademie der Wissenschaften, Boltzmannstr. 1, D-85748 Garching b. München, Germany*

**Abstract**

Intel Array Building Blocks is a high-level data-parallel programming environment designed to produce scalable and portable results on existing and upcoming multi- and many-core platforms.
We have chosen several mathematical kernels - a dense matrix-matrix multiplication, a sparse matrix-vector multiplication, a 1-D complex FFT and a conjugate gradients solver - as synthetic benchmarks and representatives of scientific codes and ported them to ArBB. This whitepaper describes the ArBB ports and presents performance and scaling measurements on the Westmere-EX based system SuperMIG at LRZ in comparison with OpenMP and MKL.

## 1. Introduction

Intel ArBB [1] is a combination of RapidMind and Intel Ct ("C for Throughput Computing"), a former research project started in 2007 by Intel to ease the programming of its future multi-core processors. RapidMind was a multi-core development platform which allowed the user to write portable code that was able to run on multi-core CPUs both from Intel and AMD as well as on hardware accelerators like GPGPUs from NVIDIA and AMD or the CELL processor. The platform was developed by RapidMind Inc., a company that started in 2004 based on the research related to the Sh project [2] at the University of Waterloo. Intel acquired RapidMind Inc. in August 2009 and combined the advantages of RapidMind with Intel's Ct technology into a successor named "Intel Array Building Blocks". A first beta version of Intel ArBB for Linux was released in September 2010. The beta program completed in November 2011 and Intel announced that they decided to transition ArBB to Intel's research and exploration portal at *http://whatif.intel.com* to broaden exposure of the technology to a wider audience.

During the PRACE preparatory phase twelve programming languages and paradigms were analysed with respect to their performance and programmability within PRACE WP 6.6 [3]. For this analysis three synthetic benchmarks from the EuroBen benchmark suite [4] have been selected. These benchmarks were a dense matrix-matrix multiplication (mod2am), a sparse matrix-vector multiplication (mod2as) and a one-dimensional complex Fast Fourier Transformation (FFT). LRZ was responsible for the RapidMind port of these three kernels [5]. We also used the kernels for synthetic performance modelling [6]. After the acquisition of RapidMind by Intel we concentrated on ArBB and ported the three EuroBen kernels and further some linear solvers (conjugate gradients, Gauss-Seidel & Jacobi solver) to ArBB.

As one of the first institutions world-wide, LRZ got early access to Intel's new "Many Integrated Core" (MIC) architecture. First experiences with the development platform on the prototype machine codenamed "Knights Ferry" have been presented at the ISC 2011 [7, 8], including some EuroBen ports. We were especially interested in ArBB as a programming model for the MIC platform. Unfortunately, due to NDA restrictions, ArBB results on the MIC platform cannot be presented in this whitepaper.

---

[*] *E-mail address*: weinberg@lrz.de.

## 2. ArBB programming model

Intel ArBB provides a C++ library interface that works with standard-conformant C++ compilers and enables the user to write portable C++ code without the need to care about the low-level parallelisation strategies of the underlying hardware architecture. The ArBB API uses standard C++ features like templates and operator overloading to create new parallel collection objects representing vectors and matrices. Further a wide variety of special operators for e.g. element-wise operations, vector-scalar operations, collectives and permutations are defined. Control flow structures mimicking C/C++ control flow are also provided. Algorithms can be expressed using mathematical notation and serial semantics. Closures can be used to capture computations for later optimisation. At compile time an intermediate representation of the code is generated which is optimised for the target architecture detected at runtime by a JIT compiler.

## 3. ArBB ports

In this whitepaper we discuss the ArBB ports of the following kernels:

- a dense matrix-matrix multiplication (EuroBen: mod2am),
- a sparse matrix-vector multiplication (EuroBen: mod2as),
- a 1-D complex Fast Fourier Transformation (FFT) (EuroBen: mod2f),
- a conjugate gradients solver for sparse linear systems.

The three EuroBen kernels are representatives of three (dense linear algebra, sparse linear algebra and spectral methods) of the so called "seven dwarves", a classification of scientific codes introduced in [9].

Performance measurements are done on the IBM BladeCenter HX5 based migration system SuperMIG at LRZ. SuperMIG will be integrated as a fat node island in the upcoming supercomputer SuperMUC at LRZ. One node of SuperMIG consists of 4 Intel Xeon Westmere-EX (E7-4870) sockets with 10 cores per socket. One Westmere-EX core running at 2.4 GHz has a double-precision peak performance of 9.6 GFlop/s. A whole node with 40 cores delivers a double precision peak performance of 384 GFlop/s and offers 256 GB of shared memory. All measurements presented in this paper use double precision arithmetic. For performance measurements of the EuroBen kernels reference input data sets as defined by [3] have been used.

ArBB supports two different optimisation levels, which can be specified at run-time by setting the environment variable `ARBB_OPT_LEVEL` to `O2` for vectorisation on a single core or to `O3` for vectorisation and usage of multiple cores. To analyse the scaling of ArBB code, the application must be linked with the development version of the ArBB library (`libarbb_dev.so`). The environment variable `ARBB_NUM_CORES` can then be used to specify the number of threads used when `ARBB_OPT_LEVEL` is set to `O3`. Hyperthreading is activated on SuperMIG, but the ArBB performance of the selected kernels cannot be improved by using more threads than physical cores. Thus we present scaling results using up to 40 threads. For performance measurements ArBB version 1.0.0.030 available under [1], Intel C++ compiler (icc) version 11.1 and MKL version 10.3 are used. The number of threads used by MKL and OpenMP is specified by `OMP_NUM_THREADS`. The environment variable `MKL_DYNAMIC` is set to FALSE, so that MKL tries not to deviate from the number of threads the user requested. To pin the threads to the cores of a node `KMP_AFFINITY="granularity=core,compact,1"` is set.

*3.1. Dense matrix-matrix multiplication (mod2am)*

The dense matrix-matrix multiplication $c = ab$ with $c_{ij} = \sum_k a_{ik} b_{kj}$ is one of most basic algorithms used in scientific computing and is also the basis of the LINPACK benchmark, which determines the TOP500 rank of a system. Various performance improvements are possible and we discuss four ArBB versions in more detail.

A naïve C implementation of the algorithm consists of three nested for-loops. An ArBB port of this naïve version is straightforward. For simplicity we assume square $n \times n$ matrices. The initialisation of the matrices and the call of the ArBB kernel function `arbb_mxm` can be implemented as:

```
1   #include <arbb.hpp>
2   using namespace arbb;
3
4   #define FLOAT double
5   #define ARBBFLOAT f64
6
7   int main() {
8     ...
9     FLOAT *a = (FLOAT*) calloc(n*n,sizeof(FLOAT));
10    FLOAT *b = (FLOAT*) calloc(n*n,sizeof(FLOAT));
11    FLOAT *c = (FLOAT*) calloc(n*n,sizeof(FLOAT));
12
13    //  initialise a,b,c ...
14
15    dense<ARBBFLOAT,2> A(n,n);
16    dense<ARBBFLOAT,2> B(n,n);
17    dense<ARBBFLOAT,2> C(n,n);
18
19    bind(A,&a[0],n,n);
20    bind(B,&b[0],n,n);
21    bind(C,&c[0],n,n);
22
23    call(arbb_mxm)(A,B,C);
24
25    C.read_only_range();
26  }
```

The basic steps consist of first including the ArBB header file `arbb.hpp` and using the C++ namespace `arbb` for simplicity (lines 1-2). ArBB defines special scalar data types like i32, f32 or f64 corresponding to the C++ data types int, float and double. (Macros for corresponding double precision floating point types are defined in lines 4-5.) Lines 9-14 allocate and initialise the $n \times n$ matrices a, b and c in regular C++ memory space. Lines 15-17 declare the two-dimensional so called "dense containers" A, B and C. Dense containers are special data types up to three dimensions used to store vectors and matrices in ArBB space. Operations that take place on such kind of dense containers are the simplest means to express parallelism in ArBB. The `bind()` function can be used to "bind" an ArBB container directly to the corresponding vector or matrix in C++ space and to automatically initialise the ArBB containers A, B and C with the values of the corresponding C++ arrays a, b and c (lines 19-21). Finally, in line 23, the ArBB kernel function `arbb_mxm` is JIT-compiled, optimised and executed via invocation of the `call()` function. Line 25 just ensures that the call has finished.

A simple ArBB port to compute the matrix elements of the naïve 3-loop version of the kernel function looks as follows:

```
1   void arbb_mxm0(dense<ARBBFLOAT, 2> a, dense<ARBBFLOAT, 2> b, dense<ARBBFLOAT, 2>& c) {
2
3     _for (usize i = 0, i != n, ++i) {
4       _for (usize j = 0, j != n;, ++j) {
5         c(i, j)= add_reduce(a.row(i) * b.col(j));
6       } _end_for;
7     } _end_for;
8   }
```

This implementation uses the special ArBB control flow constructs `_for` and `_end_for`, which are defined as macros and correspond to C++ for-loops. In line 5 the matrix element `c(i,j)` is computed by a reduction operation (sum) of a vector whose elements are defined by the scalar product of the `i`-th row of the container a and the `j`-th column of the container b.

The performance of the naïve implementation can be improved by getting rid of one of the for-loops and working directly with 2-dimensional dense containers (matrices):

```
1    void arbb_mxm1(dense<ARBBFLOAT,2> a, dense<ARBBFLOAT,2> b, dense<ARBBFLOAT,2>& c)
2    {
3    dense<ARBBFLOAT,2> t, d;
4      _for (usize i = 0, i != n, ++i) {
5        t = repeat_row(b.col(i),n);
6        d = a * t;
7        c = replace_col(c,i,add_reduce(d,0));
8      } _end_for;
```

Line 3 declares the helper matrices t and d to make the code more readable. For every iteration step with $i < n$, in line 5 the rows of the matrix t are filled with the $i$-th column of the matrix $b$, i.e. $t_{mn} = b_{ni}$. Line 6 computes the matrix $d$ by element-wise multiplication of $a$ and $t$, i.e. $d_{mn} = a_{mn}t_{mn} = a_{mn}b_{ni}$. The add_reduce(d,0) function in line 7 reduces the matrix $d$ along the 0-direction, i.e. along the rows, creating a vector $v$ with $v_m = \sum_n d_{mn} = \sum_n a_{mn}b_{ni}$. Finally, the $i$-th column of the matrix $c$ is replaced by this vector $v$, i.e. $c_{mi} = v_m = \sum_n d_{mn} = \sum_n a_{mn}b_{ni}$.

Another implementation just acts on 2-dimensional objects created by the repeat_col and repeat_row functions, without using the add_reduce() call:

```
1    void arbb_mxm2a(const dense<ARBBFLOAT, 2>& a, const dense<ARBBFLOAT, 2>& b,
2                    dense<ARBBFLOAT, 2>& c){
3
4      c = repeat_col(a.col(0), n) * repeat_row(b.row(0),n);
5      _for (usize i = 1, i < n, ++i) {
6        c += repeat_col(a.col(i), n) * repeat_row(b.row(i),n);
7      } _end_for;
8    }
```

After the initialisation of the matrix c in line 4, for each iteration step $i < n$, repeat_col(a.col(i),n) creates a matrix $\tilde{a}$ with $\tilde{a}_{mn} = a_{mi}$, and repeat_row(b.row(i),n) generates a matrix $\tilde{b}$ with $\tilde{b}_{mn} = b_{in}$. Element-wise multiplication of $\tilde{a}$ with $\tilde{b}$ yields a matrix $\tilde{c}$ with $\tilde{c}_{mn} = a_{mi}b_{in}$, which is added to the matrix $c$ in line 6.

This version was further optimised by Intel by inserting a regular C++ for-loop within the ArBB _for loop:

```
1    void arbb_mxm2b(const dense<ARBBFLOAT, 2>& a, const dense<ARBBFLOAT, 2>& b,
2                    dense<ARBBFLOAT, 2>& c) {
3
4      struct local {
5        static void mxm(std::size_t u, dense<ARBBFLOAT, 2>& c,
6          const dense<ARBBFLOAT, 2>& a, const dense<ARBBFLOAT, 2>& b)
7        {
8          c = repeat_col(a.col(0), n) * repeat_row(b.row(0), n);
9          for (std::size_t j = 1; j < u; ++j) {
10           c += repeat_col(a.col(j), n) * repeat_row(b.row(j), n);
11         }
12         const usize size = n / u;
13         _for (usize i = 1, i < size, ++i) {
14           const usize base = i * u;
15           for (std::size_t j = 0; j != u; ++j) {
16             const usize k = base + j;
17             c += repeat_col(a.col(k), n) * repeat_row(b.row(k), n);
18           }
19         } _end_for;
20
21         _for (usize i = size * u, i < n, ++i) {
22           c += repeat_col(a.col(i), m) * repeat_row(b.row(i), m);
23         } _end_for;
24       }
25     };
26     local::mxm(8, c, a, b);
```

In this version lines 8-11 compute the initial $\tilde{c}_{mn} = a_{mi}b_{in}$ for $i < u$, the optimised lines 12-19 compute the bulk of elements $\tilde{c}_{mn}$ for $u \leq i < u * size$, where the integer $size = n/u$. The remaining iterations for $u * size \leq i < n$ are computed in lines 21-23. Lines 12-19 are optimised by inserting a regular C++ for-loop with $u$ iterations (lines 15-18) into the ArBB _for loop structure (lines 13-19). Like in RapidMind regular C++ loops are executed immediately, while the special ArBB loops are recorded to build up an intermediate symbolic representation which is fed to the JIT compiler for optimization. Mind that all loop constructs in ArBB, including the _for loop, are used to describe *serial control flow* that depends on dynamically computed data. By tuning the size of $u$ the performance of `arbb_mxm2a` could be increased by a factor of two.

In our performance measurements we use square matrices with sizes $n = 10, 20, 50, 100, 192, 200, 500, 512, 576, 1000, 1024, 2000$ and $2048$.

Fig. 1 (a) and (b) show the performance of four ArBB versions on a Westmere-EX processor in comparison with a straightforward MKL implementation using the function `cblas_dgemm` and an OpenMP implementation in double precision arithmetic as a function of the matrix size. The OpenMP version is based on a naïve 3-loop version of the kernel with a `#pragma omp parallel for` before the outermost for-loop. The left figure (a) presents the single-core performance. The naïve ArBB implementation `arbb_mxm0` only reaches maximal 9% of peak performance, the slightly improved versions `arbb_mxm1` and `arbb_mxm2a` maximal approximately 30% of peak. The performance of the most improved ArBB version `arbb_mxm2b` reaches 64% of peak. Shown is also the MKL version which runs with 94% of peak performance for large matrix sizes. The right figure (b) presents the performance measurements on 40 cores using 40 threads. In this case MKL runs with maximal 58% of the peak performance of one node, while the most improved ArBB version `arbb_mxm2b` reaches up to 11% of peak performance. Using only one thread, the ArBB code for intermediate matrix sizes is around twice as fast as the serial OMP version. Using 40 threads, OpenMP outperforms ArBB for smaller matrices. The naïve implementation `arbb_mxm0` is not parallelised by ArBB and always runs single-threaded. As an example of the improved versions Fig. 1 (c) shows the scaling of the optimised ArBB version `arbb_mxm2b` with the number of threads as specified by the environment variable `ARBB_NUM_CORES`. For the largest matrix size the code scales up to approximately 15 threads. For comparison Fig. 1 (d) presents the scaling of the OpenMP port, which scales linearly up to 40 cores for large matrix sizes.

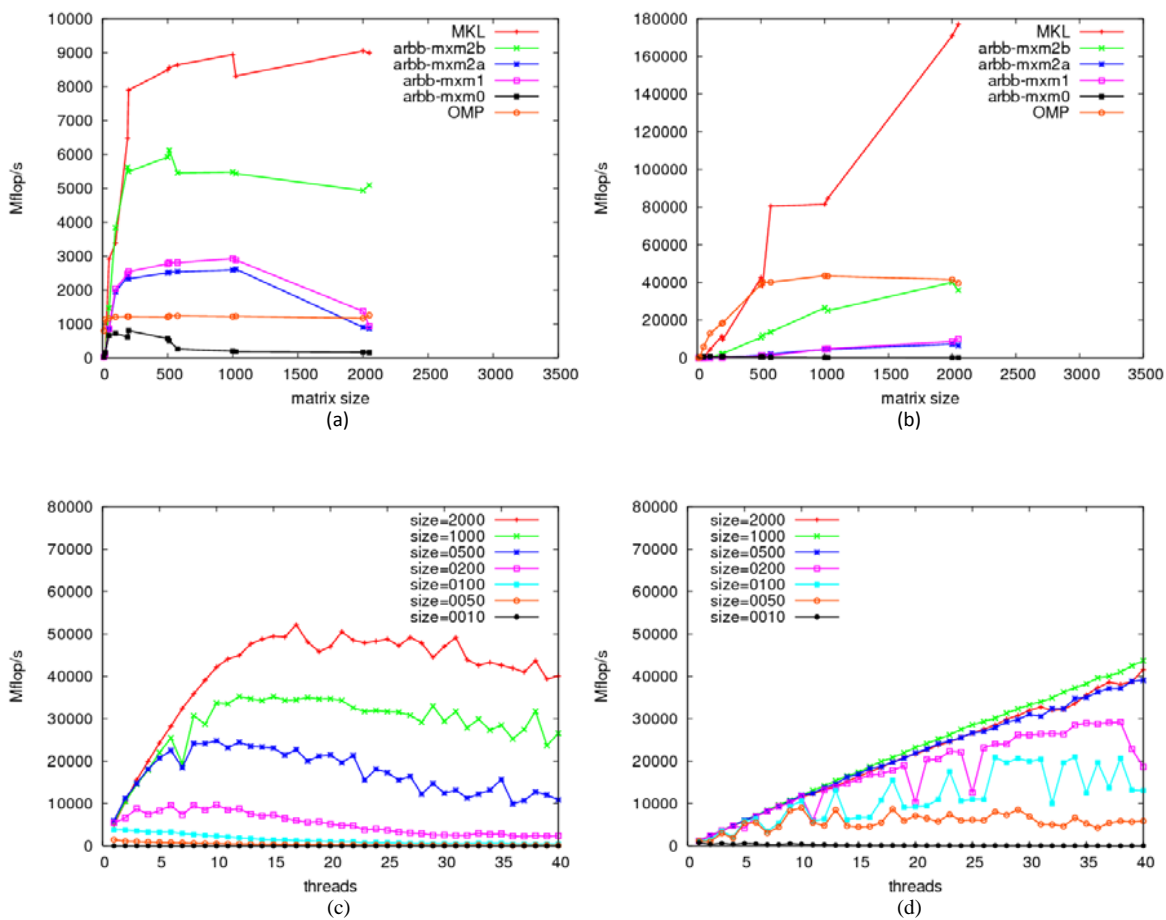

Fig. 1: Performance of various ArBB implementations of mod2am in comparison with MKL and OpenMP on a single core (a) and on a full node with 40 cores using 40 threads (b). Scaling of the optimised ArBB version `arbb_mxm2b` (c) and the OpenMP implementation (d) with the number of threads is shown for various matrix sizes.

*3.2. Sparse matrix-vector multiplication (mod2as)*

Sparse linear algebra constitutes another dwarf of the 7 dwarves of HPC. As an example of this class of algorithms the sparse matrix-vector multiplication is selected. It exposes a low computational intensity and is usually memory-bound.

The square input matrix A of mod2as is stored in a 3-array variation of the CSR (compressed sparse row) format. The array `matvals` contains the non-zero elements of A, the element i of the integer array `indx` is the number of the column in A that contains the i-th value in the `matvals` array and element j of the integer array `rowp` gives the index of the element in the `matvals` array that is the first non-zero element in row j of A.

The declaration of the input and output vectors as ArBB dense containers (lines1-6) and the call of the ArBB kernel (line 11) can be written as:

```
1    dense<ARBBINT> indx(nelmts);
2    dense<ARBBINT> rowp(nrows+1);
3    dense<ARBBFLOAT> matvals(nelmts);
4
5    dense<ARBBFLOAT> invec(ncols);
6    dense<ARBBFLOAT> outvec(nrows);
7
8    // initialise the dense objects
9    ...
10   // call ArBB kernel
11   call(arbb_spmv1)(outvec,matvals,indx,rowp,invec);
```

The implementation of the ArBB kernel function `arbb_spmv1` follows [10] and can be expressed as:

```
1    void arbb_spmv1(dense<ARBBFLOAT>& outvec, const dense<ARBBFLOAT>& matvals, const dense<ARBBINT>&
2    indx,const dense<ARBBINT>& rowp, const dense<ARBBFLOAT>& invec)
3    {
4      struct local {
5        static void reduce(ARBBFLOAT& outvec, const dense<ARBBFLOAT>& matvals, const dense<ARBBFLOAT>&
6    invec, const dense<ARBBINT>& indx, const ARBBINT& rowpi, const ARBBINT& rowpj) {
7
8          outvec = 0;
9          _for (ARBBINT i = rowpi, i != rowpj, ++i) {
10           outvec += matvals[i] * invec[indx[i]];
11         } _end_for;
12       };
13     };
14     const dense<ARBBINT> rowpi = section(rowp,0,nrows);
15     const dense<ARBBINT> rowpj = section(rowp,1,nrows);
16
17     map(local::reduce)(outvec,matvals,invec,indx,rowpi,rowpj);
18   }
```

Lines 4-13 declare the local function `reduce()` using the ArBB `_for` and `_end_for` flow control macros. The function loops over one row of the input matrix and computes the matrix-vector product which is stored in `outvec`.

`rowpi` and `ropwj` (lines 14-15) contain both `nrows` elements of `rowp`, `rowpi` starting with `rowp[0]`, and `rowpj` starting with `rowp[1]`. The ArBB `map()` function, which can only occur within a function passed to the `call()` function, permits to invoke a function written in terms of scalars across all elements of one or more dense containers. In line 17 the function `reduce()` is "mapped" to all `nrows` elements of `outvec`, `rowpi` and `rowpj`.

The performance of this kernel can be improved (`arbb_spmv2`) for sparse matrices with partly contiguous non-zero elements by distinguishing contiguous and non-contiguous sections of the input matrix and replacing line 10 in the `reduce()` function for contiguous parts of the input matrix by

```
result += values[i++] * invec[k++];
```

For the comparison with OpenMP we use two different implementations written by Filippo Spiga (CINECA) within PRACE-PP.

OMP1 mainly consists of the following parallel for-loop:

```
1    #pragma omp for private(j) schedule(runtime)
2    for( i = 0; i < nrows -1; i++ ){
3       for( j = rowp[i]; j < rowp[i+1] ; j ++ ){
4          outvec[i] = outvec[i] + matvals[j] * invec[indx[j]];
5       }
6    }
```

The improved version OMP2 is implemented as:

```
1    #pragma omp parallel  for private(t,j) schedule(runtime)
2    for( i = 0; i < nrows - 1; i++ ){
3       start_idx = rowp[i];
4       stop_idx = rowp[i+1];
5       t = outvec[i];
6       for( j = start_idx; j < stop_idx ; j ++ ){
7          t = t + matvals[j] * invec[indx[j]];
8       }
9       outvec[i] = t;
10   }
```

In our performance measurements the following $n \times n$ sparse matrices with the specified percentage of non-zero elements are considered:

| n | fill in % |
|---|---|
| 100 | 3.50 |
| 200 | 3.75 |
| 256 | 5.0 |
| 400 | 4.38 |
| 500 | 5.00 |
| 512 | 4.00 |
| 960 | 4.50 |
| 1000 | 5.00 |
| 1024 | 5.50 |
| 2000 | 7.50 |
| 4096 | 3.50 |
| 4992 | 4.00 |
| 5000 | 4.00 |
| 9984 | 4.50 |
| 10000 | 5.00 |
| 10240 | 5.72 |

Table 1: Input parameters for mod2as.

Fig. 2 (a) and (b) show the performance of the two ArBB versions as a function of the matrix size in comparison with an OpenMP port and an MKL version that uses the function `mkl_dcsrmv`. Again, the left figure (a) presents the single-core performance and the right figure (b) the performance on 40 cores using 40 threads. The MKL version only runs with maximal 16% of peak performance on a single core, the ArBB ports `arbb_spmv1` and `arbb_spmv2` reach 2% and 5%, respectively. For the selected input parameters the version `arbb_spmv2` yields better performance. Comparing with the OpenMP ports, one can see that ArBB behaves like OMP1, while MKL performs like the improved OpenMP version OMP2 when using only 1 thread. For 40 threads and large matrix sizes OpenMP and MKL perform very similarly, while the ArBB performance is much lower. On a full node using 40 threads MKL reaches 2% of peak, while the ArBB ports run with maximal 0.7% of peak performance. Fig. 2 (c) shows the scaling of the ArBB version `arbb_spmv2` with the number of threads. For the largest matrix size the code scales up to approximately 30 threads. Using more than 30 threads worsens the performance. Fig. 2 (d) presents the scaling of the OpenMP port OMP2 for comparison.

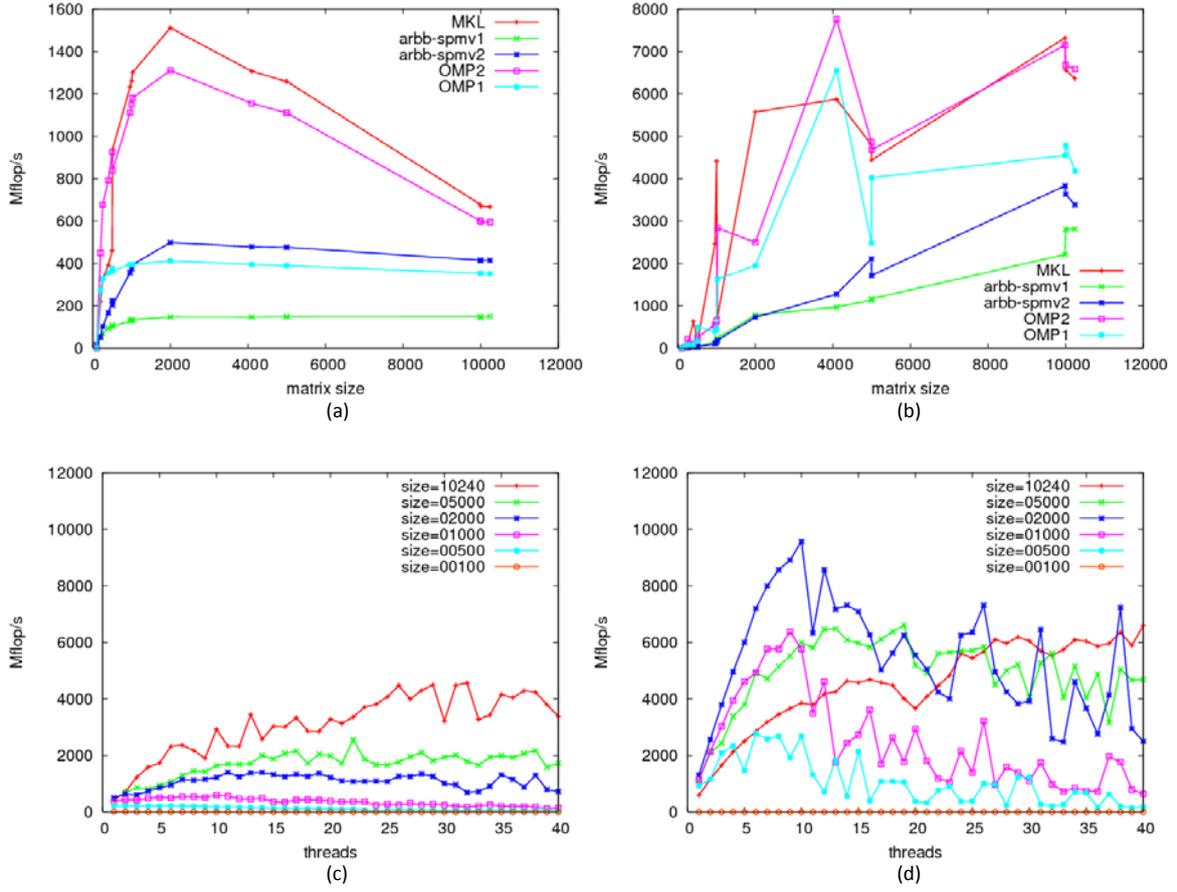

Fig. 2: Performance of two ArBB implementations of mod2as in comparison with MKL and OpenMP on a single core (a) and on a full node with 40 cores using 40 threads (b). Scaling of the ArBB version `arbb_spmv2` (c) and the OpenMP version OMP2 (d) with the number of threads is shown for various matrix sizes.

*3.3. 1-D complex Fast Fourier Transformation (FFT)*

Spectral methods build up yet another dwarf of the 7 dwarves of HPC. A widely used algorithm of this class is the Fast Fourier Transformation (FFT). To compute the 1-D complex discrete Fourier transformation (DFT)

$$F(k) = F_N(k,f) = \sum_{n=0}^{N-1} f(n)\, e^{-\frac{2\pi i k n}{N}}, \qquad (1)$$

the decimation in frequency radix-2 variant of the Cooley-Tukey algorithm is used.
The Cooley-Tukey algorithm is a well-known FFT technique that recursively breaks down a DFT of size N into smaller DFTs. The decimation in frequency variant of the algorithm divides a DFT into even/odd-numbered frequencies k. Raidx-2 means that the DFT is divided into 2 FFTs of size N/2 at each recursion level:

$$F_N(k,f) = \begin{cases} F_{N/2}\left(\frac{k}{2}, f_e\right) & \text{for k even}, \\ F_{N/2}\left(\frac{k-1}{2}, f_o\right) & \text{for k odd}, \end{cases} \qquad (2)$$

with

$$\begin{aligned} f_e(n) &= f(n) + f\left(n + \frac{N}{2}\right), \\ f_o(n) &= \left(f(n) - f\left(n + \frac{N}{2}\right)\right) e^{-\frac{2\pi i n}{N}}. \end{aligned} \qquad (3)$$

The exponential factor in (3) is called "twiddle factor". The last 2 operations together are referred to as "FFT butterfly kernel" and can be depictured as shown in Fig. 3.

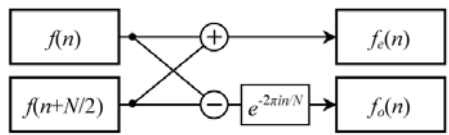

Fig. 3 : The butterfly-kernel (from [11]).

To port the Cooley-Tukey algorithm in a data-parallel manner the split-stream version developed in [11] was used. This algorithm has been designed to efficiently map the FFT computation to the stream architecture of GPUs. This version of the algorithm has already been adopted by RapidMind Inc. and is also used in example files delivered with ArBB. The advantages of this algorithm can be seen in Fig. 4.

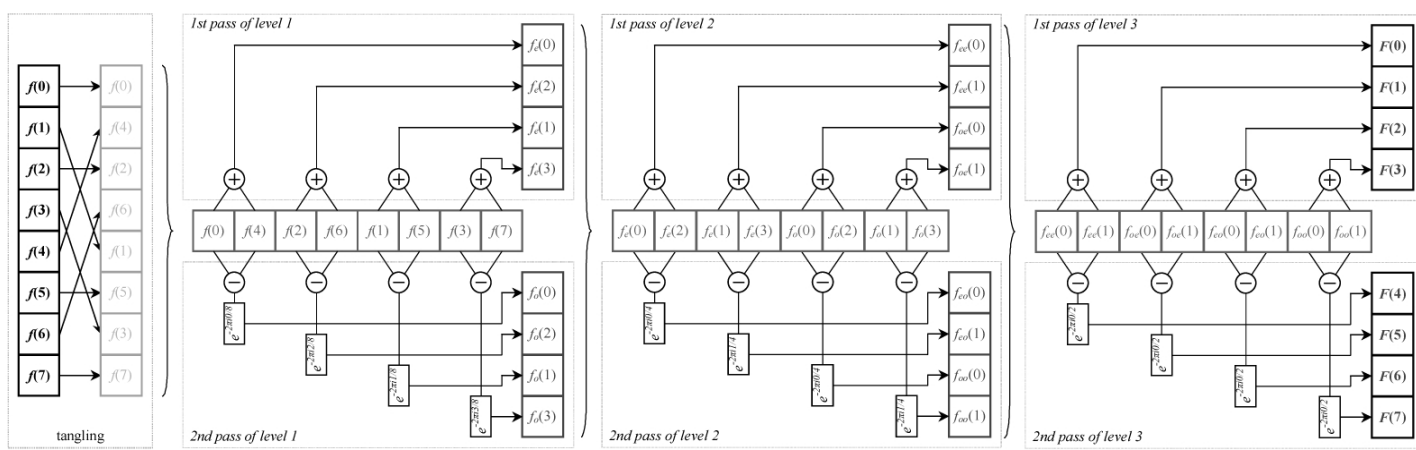

Fig. 4: Implementation of the split-stream-FFT developed by Jansen et al. (from [11]).

Only in the very beginning a reordering ("tangling") of the input data is necessary. In every iteration step the concatenation of the upper and lower output streams is the input of the next recursion level. No reordering of the output stream is necessary. Further the same operations are performed in each recursion step (except for the exact numerical values of the twiddle factors). The calculation of the butterfly kernel is split in 2 passes, one calculating $f_e(n)$ (the "up" part), the other one calculating $f_o(n)$ (the "down" part).

Ignoring the initial tangling operation of the input array, the ArBB implementation for every FFT step looks as follows:

```
1    _for (u32 i = 1, i < n, i <<= 1) {
2      dense<std::complex<ARBBFLOAT>> even = section(data,0,n/2,2);
3      dense<std::complex<ARBBFLOAT>> odd = section(data,1,n/2,2);
4
5      dense<std::complex<ARBBFLOAT>> up = even + odd;
6      dense<std::complex<ARBBFLOAT>> down = (even - odd) * repeat(section(twiddles,0,m),i);
7
8      data = cat(up,down);
9
10     m >>= 1;
11   } _end_for;
```

For each FFT step lines 2-3 gather the even and odd input streams of the input array `data`, lines 5-6 compute the up and down stream arrays. The down stream array in line 6 is multiplied with the twiddle factors, which are contained in the array container `twiddles`. Line 8 creates the output array by concatenating the up and down streams. Line 10 halves the twiddle factor sizes for the following FFT step.

The following data sizes n are used in our performance measurements: n = 256, 512, 1024, 2048, 4096, 8192, 16384, 32768, 65536, 131072, 262144, 524288 and 1048576. Fig. 5 (a) shows the single-core performance of the ArBB port as a function of the FFT data size in comparison with MKL and two different serial versions. MKL uses the DftiComputeForward routine. For the serial version we compare a simple radix-2 Cooley-Tukey implementation with a serial split-stream implementation and an optimised combined radix-4 and radix-2 implementation from the EuroBen suite (CFFT4).

The MKL version runs with maximal 31% of peak performance. The maximal performance of the ArBB port is 5% of peak, which is comparable to the simple serial radix-2 version with 3% of peak, but much below the optimised serial CFFT4 code from the EuroBen suite with 17 % of peak. Fig. 5 (b) presents the scaling of the ArBB port with the number of threads. Except for the largest data size the performance steeply drops with increasing number of threads.

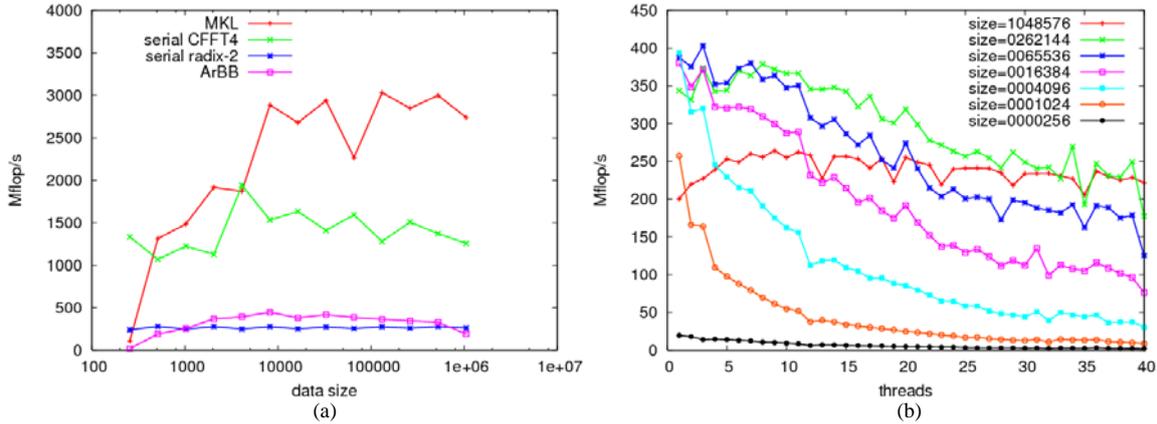

Fig. 5: (a) Performance of the ArBB implementations of mod2f in comparison with MKL and two serial implementations on a single core as a function of the data size. (b) Scaling of the ArBB port of mod2f with the number of threads for various data sizes.

*3.4. Conjugate gradients solver for sparse linear systems*

The conjugate gradients algorithm is widely used for the numerical solution of linear equations with symmetric and positive-definite matrices. As an iterative method it is well suited for sparse linear systems that are too large to be handled by direct methods. The textbook version [12] of the conjugate gradients algorithm reads as follows:

---

If $A \in \mathbb{R}^{n \times n}$ is a symmetric positive definite matrix, $b \in \mathbb{R}^n$, and $x_0 \in \mathbb{R}^n$ is an initial guess, then the following algorithm computes $x \in \mathbb{R}^n$ so $A x = b$.

**Initialisation**:
- $x_0$ = initial guess, e.g. $x_0 = 0$
- $r_0 = b$ (random vector)
- $p_0 = b - A x_0$

**Iteration**: (while $|r_n|^2 >$ stop)
- $x_{n+1} = x_n + \alpha_n p_n$
- $r_{n+1} = r_n - \alpha_n A p_n$ («residual»)
- $p_{n+1} = r_{n+1} + \beta_n p_n$ («search direction»)
- $\alpha_n = \frac{(r_n, r_n)}{(p_n, A p_n)}$
- $\beta_n = \frac{(r_{n+1}, r_{n+1})}{(r_n, r_n)}$

---

Fig. 6: Textbook version of the conjugate gradients algorithm (from [12]).

This algorithm can almost be literally rewritten in ArBB syntax:

```
1    i32 k=0;
2    ARBBFLOAT r2 = add_reduce(b*b);
3    _while(r2 > stop  && (k < max_iters))
4    {
5      // Compute sparse matrix-vector multiplication Ap=A*p
6      arbb_spmv(Ap, csrVals, csrColPtr, csrRowPtr, p);
7    
8      alpha  = r2 / add_reduce(p*Ap);
9    
10     r2_old = r2;
11     r = r-alpha*Ap;
12     r2 = add_reduce(r*r);
13   
14     beta= r2 / r2_old;
15   
16     x = x+alpha*p;
17     p = r+beta*p;
18   
19     ++k;
20   } _end_while;
21  }
```

This example demonstrates the elegance of the simple, math-like notation of ArBB. We use banded symmetric $n \times n$ matrices ($n$ = 128, 256, 512 and 1024) with bandwidths $bw$ between 3 and 511 as sparse matrices, which are again stored in CSR format.

The following input parameters are used:

| # conf | n | bw |
|---|---|---|
| 1 | 128 | 3 |
| 2 | 128 | 31 |
| 3 | 128 | 63 |
| 4 | 256 | 3 |
| 5 | 256 | 31 |
| 6 | 256 | 63 |
| 7 | 256 | 127 |
| 8 | 512 | 3 |
| 9 | 512 | 31 |
| 10 | 512 | 63 |
| 11 | 512 | 127 |
| 12 | 512 | 255 |
| 13 | 1024 | 3 |
| 14 | 1024 | 31 |
| 15 | 1024 | 63 |
| 16 | 1024 | 127 |
| 17 | 1024 | 255 |
| 18 | 1024 | 511 |

Table 2: Input parameters for the conjugate gradients solver.

For the sparse matrix-vector multiplication in line 6 the routines `arbb_spmv1` or `arbb_spmv2` from mod2as are used for the ArBB ports. Fig. 7 (a) shows the single-core ArBB performance for various banded sparse matrices as a function of the configuration number in comparison with a simple serial version and a version using the MKL function `mkl_dcsrmv` for the sparse matrix-vector multiplication. For larger bandwidths the ArBB version calling `arbb_spmv2` is faster than the one calling `arbb_spmv1`, as expected. However, both versions are much slower than the serial version. Calling the MKL version out of the serial version is only beneficial for larger matrices with larger bandwidths. Fig. 7 (b) presents the scaling behaviour of the ArBB implementation using `arbb_spmv2` for $n$ =1024 and various bandwidths (#conf=13-18). For larger bandwidths the code scales up to 7 threads, while in other cases the performance drops with increasing number of threads.

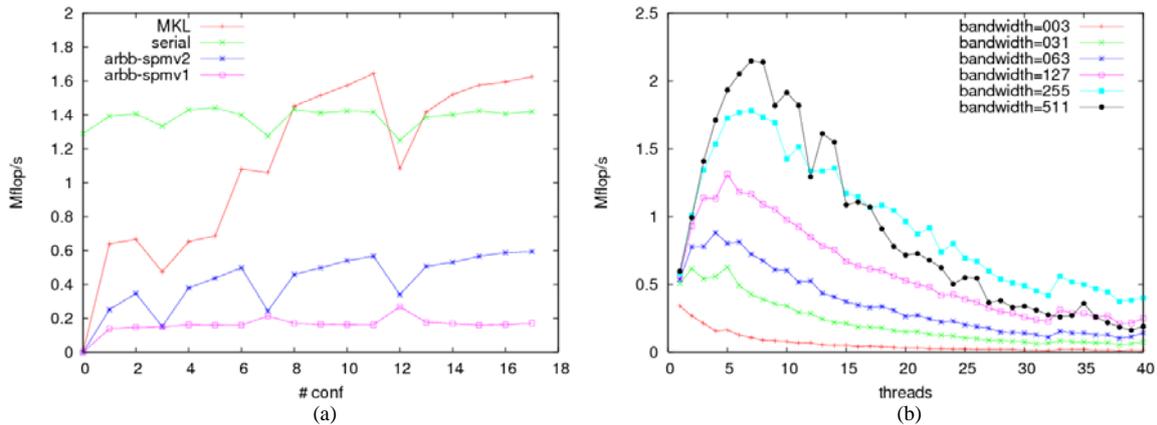

Figure 7: (a) Performance of the ArBB implementations of the conjugate gradient solver for sparse banded matrices in comparison with a serial version and a version calling MKL for the sparse matrix-vector multiplication. The performance is shown as a function of the configuration number. (b) shows the scaling behaviour of the ArBB implementation using `arbb_spmv2` for $n$ =1024 and various bandwidths (covering #conf=13-18).

## 4. Summary

ArBB is a powerful language for expressing data-parallelism in a simple way. The main advantages of ArBB are the availability of various operations for manipulating vectors and matrices and the simple, serial mathematic-like semantic to express parallelism. The possibility to use closures to capture computations and the run-time optimisation of the captured code by a JIT compiler are very powerful concepts. The development time using ArBB is rather low for people who are used to program in C++. A Fortran interface is not supported. ArBB is limited to shared memory systems, since internally the pthreads, OpenMP and TBB libraries are used. The distinction of C++ and ArBB memory space and the definition of incompatible corresponding data types lead to some overhead in the code and complicate the parallelisation of existing code. ArBB is currently restricted to x86 based systems and (under NDA) the Intel MIC architecture. Compared to the former RapidMind product, which supported many- and multi-core CPUs, GPGPUs and the Cell processor, the portability of ArBB is quite limited.

The performance and the scaling of code compiled with the current version of ArBB is partly still rather poor. In the case of the matrix-matrix multiplication mod2am the performance of simple data-parallel implementations reaches approximately 30% of peak performance (double precision) on a Westmere-EX core. The performance of mod2am could be improved by a factor of two with support by Intel by loop restructuring, but we would expect the runtime optimiser to establish such reconstructions rather than the programmer. For large data sets the codes mod2am and mod2as scale up to approximately 15 and 30 cores, respectively, on a Westmere-EX node with 40 cores. In other cases, especially for the FFT, scaling is insufficient. However, it should be pointed out that ArBB has been mainly developed as a new data-parallel programming language and not as a highly optimised numerical library like MKL. Intel decided not to merchandise ArBB as a product in near future. We hope that the great work done by the RapidMind and ArBB developers will influence future parallel programming languages and products.

## 5. Acknowledgements

The author thanks Hans Pabst (Intel) for insightful discussions and very constructive support in optimising the codes mod2am and mod2as. This work was financially supported by the PRACE project funded in part by the EUs 7th Framework Programme (FP7/2007-2013) under grant agreement no. RI-211528 and FP7-261557 and the KONWIHR-II project "OMI4papps: Optimisation, Modelling and Implementation for Highly Parallel Applications".